\documentclass{jpc} 

\usepackage{makecell}
\usepackage{tabularx}

\usepackage{color, colortbl}
\usepackage{soul}
\definecolor{Gray}{gray}{0.9}

\usepackage{algpseudocode}

\usepackage{boldline}
\graphicspath{{./images//}}

\newcolumntype{L}[1]{>{\raggedright\let\newline\\\arraybackslash\hspace{0pt}}m{#1}}
\newcolumntype{C}[1]{>{\centering\let\newline\\\arraybackslash\hspace{0pt}}m{#1}}
\newcolumntype{R}[1]{>{\raggedleft\let\newline\\\arraybackslash\hspace{0pt}}m{#1}}


\usepackage{natbib}
\usepackage[ruled]{algorithm2e}
\theoremstyle{plain} 

\begin{document}

\title[Incompatibilities in Current Practice]{Incompatibilities Between Current Practices in Statistical Data Analysis and Differential Privacy}

\author[J.~Snoke]{Joshua Snoke}	
\address{RAND Corporation, Pittsburgh, PA, USA}	
\email{jsnoke@rand.org}  

\author[C.M.~Bowen]{Claire McKay Bowen}	
\address{Urban Institute, Washington D.C., USA}	
\email{cbowen@urban.org}  

\author[A.R.~Williams]{Aaron R. Williams}	
\address{Urban Institute, Washington D.C., USA}	
\email{awilliams@urban.org}  

\author[A.F.~Barrientos]{Andr\'es F. Barrientos}	
\address{Department of Statistics, Florida State University, Tallahassee, FL, USA}	
\email{abarrientos@fsu.edu}  




\begin{abstract}
  \noindent The authors discuss their experience applying differential privacy with a complex data set with the goal of enabling standard approaches to statistical data analysis. They highlight lessons learned and roadblocks encountered, distilling them into incompatibilities between current practices in statistical data analysis and differential privacy that go beyond issues which can be solved with a noisy measurements file. The authors discuss how overcoming these incompatibilities require compromise and a change in either our approach to statistical data analysis or differential privacy that should be addressed head-on.
\end{abstract}

\maketitle

\section{A Brief History of Differential Privacy in the Wild}
Researchers and data practitioners make many different claims concerning differential privacy (DP) and its impact on statistical analysis. Some maintain that DP provides the future for how government agencies and private companies will release public statistics and data sets. Others argue that pursuing DP is a mistake and will destroy how we disseminate information as we know it. These debates often center on notions of the trade-off between accuracy and utility, selecting privacy budgets, or the appropriate definition of privacy loss. While these questions matter, they often fail to acknowledge other underlying issues. While DP is a framework containing a wide variety of implementations, there exist fundamental incompatibilities between standard statistical analysis approaches and the possibilities within a DP framework\footnote{One might rightfully respond that some incompatibilities are not unique to DP and also exist for alternative statistical disclosure control methods, and we would agree. For the sake of simplicity, we do not dig into those questions in this piece. See \cite{slavkovic2023statistical} for a thorough article on the similarities and differences between DP and other statistical disclosure control approaches.}.

From the position of either a statistical analyst or a privacy practitioner, dealing with these incompatibilities often comes across in the field as claims that the other side is ``doing it wrong." In reality, we have two paradigms functioning under different assumed rules and making them compatible will require some changes to one or both frameworks. We write this perspective from the point of view of both privacy researchers and statistical analysts who are broadly approaching their analysis from a frequentist perspective with the goal of statistical inference. We also use the term statistical analyst as a catch-all term encompassing statisticians, economists, demographers, or any data scientist working in social statistics.

Briefly considering the history of practical DP implementations, early applications of the framework generally obscured the incompatibilities due to the specific use-cases. For instance, numerous tech companies created interactive or query-based DP frameworks that allowed analysts to submit a question and receive a noisy statistic in return. Some examples include audience engagement statistics on LinkedIn \citep{rogers2021linkedin}, SQL queries in Uber's driver and rider database \citep{johnson2018towards}, people's movement on Google Maps within certain geographic regions \citep{aktay2020google}, and HealthKit usage on Apple products.\footnote{Differential Privacy Team, Apple. n.d. ``Learning with Privacy at Scale.'' https://docsassets.developer.apple.com/ml-research/papers/learning-with-privacy-at-scale.pdf.} These applications enable analysts to perform learning tasks or other tasks that do not rely on uncertainty estimates or explicit hypothesis testing, since statistical learning was the goal for the use-case.

When the U.S. Census Bureau announced their adoption of DP for the 2020 Census data products, it significantly increased the debate between DP advocates and DP skeptics of its applicability to demographic data. At a high level, the 2020 Disclosure Avoidance System added differentially private noise to thousands of statistics at different census geographic level and corrected for any inconsistencies (e.g., ensuring that population counts for census tracts sum to the county count) using a complex post-processing method formulated in the TopDown Algorithm. This change in statistical data privacy protection created huge backlash from the data user community, including lawsuits\footnote{``Alabama drops lawsuit challenging Census privacy method,'' https://apnews.com/article/alabama-lawsuitscensus-2020-redistricting-us-census-bureau-3c6f5eacc6c5638756700ba8308c45d2} and a letter to the Director of the U.S. Census Bureau\footnote{``Researchers ask Census to stop controversial privacy method
,'' https://apnews.com/article/census-2020-us-bureau-government-and-politics-20e683c71eeb62ee4b7792d7d8530419} to stop the use of DP on the 2020 Census data products. While some simply urged not to add noise, constructive critiques pointed out that statistical analysts needed the ability to adjust for the noise in order to conduct their usual statistical analyses. In essence, analysts needed new tools that are not part of traditional statistical data analysis. The introduction of bias or increased uncertainty in estimates due to privacy protections was not new to those who had been working in statistical disclosure control. What was novel is that DP promised to account or track the added noise rather than ignoring it. But, for some time, it was not clear how this promise would be achieved.

In response, several leading researchers requested the Census Bureau release the noisy measurements data set \citep{dwork2021letter}. The researchers' reasoning was that access to the noisy measurement file (NMF) would allow analysts to account for the noise introduced from the TopDown Algorithm. In addition, even when statistical analysts have access to the noisy measurement file, most do not have the background understanding of DP or the computational ability to make the required corrections. Given this challenge, some privacy researchers organized a workshop on the ``Analysis of Census Noisy Measurement Files and Differential Privacy.'' The purpose of the workshop was to convene experts from various fields and practices within social science, demography, public policy, statistics, and computer science to discuss the need for implementable tools that allowed analyses on privacy preserving noise induced data and statistics.\footnote{``Workshop on the Analysis of Census Noisy Measurement Files and Differential Privacy,'' Accessed February 14, 2023. http://dimacs.rutgers.edu/events/details?eID=2038} 

The debate concerning the decennial Census and the noisy measurement file helped illuminate the distinction between the way noise is added and the ability to account for it in statistical analyses. Some argued that reverting to previous statistical data privacy methods, such as swapping, would be preferable to DP. But fundamentally these methods have the same issues as the lack of a NMF and using them would not provide statistical analysts with any better means of analyzing the data. Without a means of accounting for the noise introduced to protect privacy, any statistical analyses will be biased or falsely overpowered.

\section{Differentially Private Query Systems for Statistical Inference}
At this point in the story, it may be tempting to argue that any issues performing statistical analysis under DP can be handled by the development of NMFs and tools to account for the additional noise. The widespread publicity of the Census Bureau's adoption of DP locked in, for many, a particular means of achieving DP, with the corresponding solution being access to the NMF. But these solutions only makes sense under the scenario where a set of predetermined measurements are made and released non-interactively, such as publishing a public data set. In fact, the example of the decennial Census is still \textit{rather similar} to applications of DP in the tech space. The underlying data were very large and the queries were counts. 

Implementing DP on other products, such as smaller surveys or interactive query systems with the purpose of statistical inference, introduces an entirely new set of incompatibilities. The adoption of DP for settings where researchers make sequences of queries that encompass more complex statistical analyses will require the field to overcome more significant barriers than those that faced the Census Bureau.

In our work, \cite{barrientos2021feasibility}, we explore creating a differentially private query system, known as a validation server, to allow tax researchers to estimate simple statistics and linear regressions on confidential IRS data. In contrast to the decennial census, this system must allow for interactive queries and include statistic-specific uncertainty estimates for hypothesis testing, such as confidence intervals.

When conducting that study, the first difficulty we encountered was that only a small fraction of the published papers proposing DP algorithms for querying common statistics, such as means and regression coefficients, provide uncertainty estimates. Another challenge was that accurate queries required users to input substantial information about the distribution or the range of the underlying data which is not commonly known. Additionally, an astute user needs to determine how to allocate their privacy budget. But, in many cases, they may not know the complete set of queries they want to perform before starting. Finally, designing an interactive query system that allows different types of queries with optimal performance under DP is not guaranteed to compose in a trivial manner, though it is theoretically possible \cite{rogers2016privacy,whitehouse2023fully}.

While still in progress, working with an interdisciplinary team has pushed us towards considering compromises that may dissatisfy either the data analyst or the privacy practitioner. Under one of our proposed systems, users may develop and test analyses on non-formally private synthetic data that represent a more limited subset of individuals prior to submitting their analysis to the validation server. Conversely, our server includes a much more limited set of allowed queries than a typical tax economist would expect. In particular, methods for working with data from complex surveys that need to incorporate survey weights, the backbone of the federal statistical system, do not exist.

Another proposed compromise is using two sets of privacy budgets, such that users can conduct some initial analyses prior to exhausting their budget on estimates that can be released. On the other hand, users still carry the burden of bringing prior information for their analyses, such as indicating the range of possible values or spending some privacy budget to estimate it. At this point, we do not know the full implications of these decisions or whether they will be featured in the eventual implementation of the validation server. We only highlight the various compromises we have wrestled through to enable statistical analysis and DP to function together.

\section{Incompatibilities That Require Compromise}

Based on the lessons we have learned, we offer the following general incompatibilities between DP and normal statistical practice that must be addressed in practice implementations. When the goal is traditional statistical analysis, specifically inference tasks, overcoming the incompatibilities requires compromises, either from the statistical analysts, formal privacy practitioners, or both.

(a)	\textbf{Estimates for traditional statistical inference.} Frequentist methods for statistical inference rely on estimates, such as confidence intervals or $p$-values to perform hypothesis testing. DP methods have only been shown to provide guarantees for statistical accuracy in scenarios where the size of the data set is large enough \citep[e.g.,][]{chaudhuri2011differentially,sheffet2017differentially,pena2021differentially,sart2023density}, whereas other papers simply choose not to evaluate statistical inference tasks. While large sample properties, also referred to as asymptotic properties, are universally desirable, they frequently fail to provide substantial insight into what can be expected for specific finite sample sizes. This is because the condition of being ``large enough'' is difficult to define. Additionally, post-processing can frequently induce bias, and DP implementations have not been shown in practice to provide estimates with amounts of noise that statistical analysts would consider reasonable without sacrificing substantial amounts of privacy\footnote{This is measured through the privacy parameter, most commonly $\epsilon$, associated with the definition of differential privacy that a method satisfies.}. Furthermore, effective DP methods do not exist for more complex models involving survey weights, panel data, or methods for causal inference.

A few compromise options exist to handle this incompatibility. First, a statistical data user can only query point estimates and decide not to perform frequentist hypothesis testing. Alternatively, users may opt for Bayesian approaches. While requiring privacy budgets similar to frequentist methods, Bayesian approaches can account for the assumptions and probabilistic nature behind DP, and they can be used for full inference about the parameters and predictions. Specifically, Bayesian techniques allow for the simultaneous consideration of various factors, such as the use of non-sufficient summary statistics, assumptions about data boundedness (clamping), and noise injection for privacy. These elements are difficult for traditional frequentist statistics to handle simultaneously. Though Bayesian methods offer a promising alternative, they reflect a compromise because they remain unused and unfamiliar to many in the statistical research community. The field needs concrete work answering this question. If we need to consider alternative approaches to frequentist inference, how will this impact statistical practice? If we can only conduct frequentist inference under different privacy definitions, is accurate statistical inference possible?

(b)	\textbf{Control or nuisance variables.} In estimating regression models, it is very common to include control variables for which the estimates are not really of interest to the person querying the model. Current differentially private methods do not account for this, nor is it clear what it would look like if they did. In regression, for example, queries that include uncertainty estimates, namely those that perturb the sufficient statistics, add noise that grows polynomially with the number of predictors. Other methods, such as \cite{wang2018revisiting}, scale better with predictors but require a Bayesian inference approach and have not been shown to be practically implementable \citep{barrientos2021feasibility}. Whatever the approach, queries which return these parameters lead to heightened noise added to estimates of interest as the number of control variables grows, and it can be argued that this noise is extraneous even if it only scales linearly.

Can DP be reformulated to ignore the privacy-loss due to these control variables? Is including control variables in a regression query but not receiving coefficients possible? How would this impact the privacy guarantee? If so, is there still value in spending privacy budget on nuisance variables? Conversely, can statistical practice be changed such that appropriate analyses can be made without including control variables?

(c)	\textbf{Assumptions on the range of the data or other assumptions.} A common barrier for statistical analysts using DP methods is the need to place prior bounds on the distribution of data or statistics in order to calculate a finite sensitivity. Knowing this information or finding ways of estimating it apart from the data is not always part of statistical practice. In many cases, there are no good priors to help set these bounds, and the DP literature is largely silent on this problem. 

In some cases, analysts may be able to estimate the bounds under DP. For example, \cite{wilson2019differentially} propose a method for automatically estimating bounds on continuous data by taking advantage of the physical limitations of machine precision and minimizing the amount of data clipping. This helps select bounds without prior information, but a fundamental bias-variance trade-off exists when privately estimating bounds \cite{amin2019bounding}. More problematically, the analysts cannot know where they fall on this bias-variance trade-off without knowing the real bounds of the data. This inhibits inferential methods to adjust for the fact that the final estimates include uncertainty both from the noise mechanism of the final query and the noise mechanism of the prior bound-setting query. Either statistical analysts will need to adapt their methodology to account for this, or DP methods will need to adapt to enable analysts to estimate the impact of privately setting bounds.

(d)	\textbf{Performing exploratory data analysis.} Statistical analysts commonly explore the data using visualizations, marginal and multivariate summary statistics, and model diagnostics. Most statistical researchers, who we assume are not trusted to access the private data, do not know ahead of time exactly what analyses will be run. In one sense, DP can help disincentivize bad exploratory data analysis (EDA) practices, such as $p$-hacking. For example, if an analyst must split their privacy budget across EDA queries, they may limit the amount of exploration they make, potentially making it less likely for them to find spurious results. Conversely, DP may make it more difficult to account for multiple testing, since the final inferences should account for the uncertainty propagated through all the analyses performed to select the final model. Given the issues discussed earlier concerning frequentist inference under DP and the lack of work on multiple testing, it is not clear whether this is feasible.

Though some in the broader scientific community are moving towards pre-specifying every model in research\footnote{For example, \url{https://plos.org/open-science/preregistration/}.}, this is far from the current reality in all disciplines. And not all EDA results in $p$-hacking. Data users need the ability to probe assumptions or look for data abnormalities, and they may run into serious problems without this ability. For example, without understanding the data a user may request for a regression model where a predictor has no variance. Under DP, this query may return random noise or a null result with some probability, and statistical analysts are not prepared for this type of response. Either statistical analysts will need to adapt their research without the ability to do EDA, or DP methods need to find ways to allow EDA in a private setting.

(e) \textbf{Limited queries and the privacy budget.} Finally, there are a broader set of issues that come from performing statistical analyses with a limited privacy budget. This is a concept both unfamiliar to statistical analysts and carrying significant implications. For example, how should data maintainers allocate budgets to numerous analysts? How do analysts determine how much of their budget to allocate to multiple model specifications, robustness checks, and the final models? What should analysts do if a journal reviewer asks for alternative model specifications or other requests, such as reproducing the results, and there is no more privacy budget to spend? 

More challenges occur when multiple data users or analysts submit the same analysis. As an illustrative example, imagine there are two data users (A and B), who submit the same analysis on the same part of the confidential data, but user A submitted before data user B. The validation server could handle this situation in two ways. 

One approach is to consider the analyses from user A and B as separate analyses. In other words, these analyses would likely produce two different results regardless if user A and B used the same or different privacy loss budgets. Under this approach, both data users would not be notified about each others' analyses, ensuring greater confidentiality and encouraging analysts to use the validation server results more confidently. Knowing their specific research ideas will not be revealed mitigates concerns about being scooped or having their ideas preempted. But producing different answers for the same query creates communication and education problems in explaining to both data users that their answers are valid. It is also means, from a societal point of view, that we are sacrificing more privacy (or accuracy) to run the same analysis twice.

The other approach is to apply the same result from data user A for data user B. Unlike the first approach, there would be no confusion of having two different results for the same analysis. The data users could also share the cost of the privacy loss budget, spending less of their individual budgets. However, both data users would be informed that the analysis was conducted twice. We could even make this more complex and extend it to the situation where data user B wanted a more accurate result and spend more privacy loss budget than data user A.

These are some of many other issues we need to address. In any case, either statistical analysts will need to develop novel means of optimally allocating their budget for their research or DP will need to rethink budgets in interactive query systems.

\section{Future Steps}
We hope this perspective will serve as a means of calling out the broader set of issues beyond problems that can be solved using a noisy measurements file. As of now, it is not clear whether the compromise comes from the way we perform statistical analyses, the way we implement DP, or both. We hope that future papers on DP will wrestle through these practical questions in a larger way than has been typically done to this point.

\section*{Acknowledgments}
This research was funded by the Alfred P. Sloan Foundation [G-2020-14024] and National Science Foundation National Center for Science and Engineering Statistics [49100420C0002 and 49100422C0008].

\bibliography{ref}
\bibliographystyle{abbrvnat}

\end{document}